\documentclass[twocolumn]{aa}
\usepackage{graphicx}

\def\sax{{\it Beppo}SAX$~$}
\def\aro{{$\alpha_{\rm ro}$~}}
\def\aox{{$\alpha_{\rm ox}$~}}

\def\amx{{$\alpha_{\mu x}$~}}
\def\amg{{$\alpha_{\mu \gamma}$~}}
\def\amgB{{$\alpha_{{\mu \gamma}_{100\%CGB}}=0.994$~}}
\def\arm{{$\alpha_{5-94GHz}$~}}
\def\ar{{$\alpha_{\rm r}$~}}
\def\ax{{$\alpha_{\rm x}$~}}
\def\ergs{{erg~cm$^{-2}$s$^{-1}$~}}
\def\ergssd{{erg~cm$^{-2}$s$^{-1}$deg$^{-1}$~}}
\def\ergj{{erg~cm$^{-2}$s$^{-1}$Jy$^{-1}$~}}
\newcommand{\lsim}{{\lower.5ex\hbox{$\; \buildrel < \over \sim \;$}}}
\newcommand{\gsim}{{\lower.5ex\hbox{$\; \buildrel > \over \sim \;$}}}

\def\fxfr{$f_{\rm x}/f_{\rm r}$~}
\def\nupeak{$\nu_{peak}$~}

\def\ee{\end{equation}}
\def\be{\begin{equation}}

\begin{document}
\title{Non-thermal Cosmic Backgrounds from Blazars: \\
the contribution to the CMB, X-ray and  $\gamma$-ray Backgrounds}
\author{P.~Giommi\inst{1,2},
 S.~Colafrancesco\inst{3},  E.~Cavazzuti\inst{1,2},  M.~Perri\inst{1} and C.~Pittori
\inst{1,4}
 \institute{
            ASI Science Data Center, ASDC c/o ESRIN,
            via G. Galilei 00044 Frascati, Italy.
\and
            Agenzia Spaziale Italiana,
            Unit\'a Osservazione dell'Universo,
            viale Liegi, 26 00198 Roma, Italy
\and
            INAF - Osservatorio Astronomico di Roma
            via Frascati 33, I-00040 Monteporzio, Italy.
\and
            Universit\'a di Roma "Tor Vergata" and INFN sez. Roma 2,
            Italy
}}
\offprints{paolo.giommi@asi.it}
\date{Received: ; Accepted: }

\authorrunning{P. Giommi et al.}
\titlerunning{Non-thermal Cosmic Backgrounds from Blazars}

\abstract{We present a new assessment of the contribution of the Blazar population to the extragalactic
background radiation across the electromagnetic spectrum.
Our calculations rely on deep Blazar radio counts that we have derived combining several
radio and multi-frequency surveys.  We show that Blazar emission integrated over cosmic time
gives rise to a considerable broad-band non-thermal Cosmic Background that in some parts of the
electromagnetic spectrum dominates the extragalactic brightness.

We confirm that Blazars are the main discrete contributors to the Cosmic Microwave Background
where we estimate that their integrated emission causes an apparent temperature increase of 5-50 $\mu$K in the frequency range 50-250 GHz. The CMB primordial fluctuation spectrum is contaminated starting at multipole $l \approx 300-600,$ in the case of  a  completely random source distribution, or at lower $l $ values if  spatial clustering is present.
We estimate that well over one hundred-thousand Blazars will produce a significant
signal in the maps of the upcoming Planck CMB anisotropy mission. Because of a tight correlation
between the microwave and the X-ray flux these sources are expected to be X-ray emitters with flux
larger than a few $ 10^{-15}$ \ergs in the soft X-ray band.
A large fraction of the foreground sources in current and near future CMB anisotropy maps could
therefore be identified and removed using a multi-frequency approach, provided that
a sufficiently deep all sky X-ray survey will become available in the near future.

We further show that Blazars are a major constituent of all high energy extragalactic backgrounds.
Their contribution is expected to be 11-12 \% at X-ray frequencies and possibly 100\% in the
$\sim 0.5-50 $ MeV band.
At higher energies ($E > 100~$MeV) the estimated Blazar collective emission, obtained extrapolating
their integrated micro-wave flux to the $\gamma$-ray band using the SED of EGRET detected
sources, over-predicts the extragalactic background by a large factor, thus implying that
Blazars not only dominate the  $\gamma$-ray sky but also that their average duty cycle
at these frequencies must be rather low.
Finally, we find that Blazars of the HBL type may produce a significant amount of flux at
TeV energies.


}

\maketitle

\section{Introduction}

The identification of the first quasar (\cite{schmidt63}) marked the beginning of AGN astrophysics but
also the discovery of the first flat spectrum radio quasar (3C273), a type of highly variable, often
polarized extragalactic radio sources that, together with the even more puzzling BL Lacertae objects,
make the class of Blazars, the most extreme type of Active Galactic Nuclei known. The Blazar typical
observational properties include the emission of electromagnetic radiation across the entire spectrum,
from radio waves to the most energetic $\gamma$-rays, irregular rapid variability, apparent
super-luminal motion, flat radio spectrum, large and variable polarization at radio and, especially,
at optical frequencies.

Blazar of the BL Lacertae type (BL Lacs) are distinguished by non-thermal emission with no (or very
weak) emission lines, are often associated with the nuclei of elliptical galaxies and are the only
population of extragalactic sources that shows negative cosmological evolution (\cite{Bade98, Gio99,
Rec00}). Flat Spectrum Radio Quasars (FSRQs) share the strong non-thermal emission of BL Lacs but also
show intense broad line emission and strong cosmological evolution similar to that of radio quiet QSOs
(\cite{Cac02, L01, Wall85}).

\begin{figure}[ht]
\vbox{
\centerline{
\includegraphics[width=6.5cm, angle=-90]{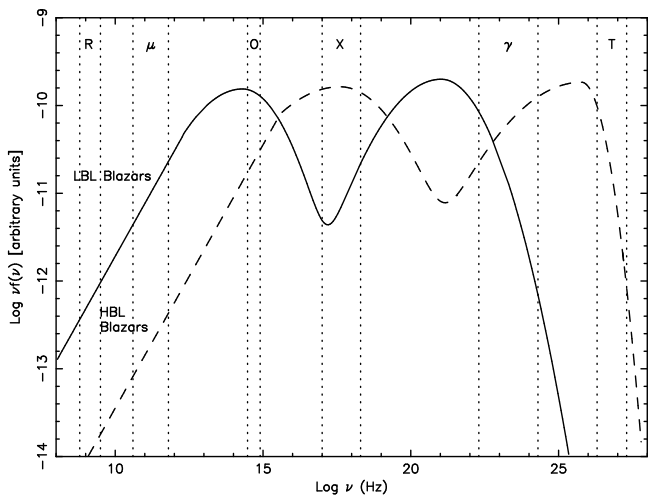}
}}
\caption{The Spectral Energy Distribution of different types of Blazars represented by
Synchrotron-Self-Compton models with emission peaking at different energies. LBL sources are those
where the synchrotron peak is located at Infra-Red frequencies and therefore their X-ray emission is due to
Inverse Compton radiation. HBL Blazars are those where the synchrotron component peaks in the
UV/X-ray band making their \fxfr flux ratio hundreds or thousands of times higher than that of LBLs.
}
\label{all_bl_sed}
\end{figure}

Blazars are widely assumed to be powerful sources emitting a continuum of electromagnetic (e.m.)
radiation from a relativistic jet viewed closely along the line of sight (\cite{bla78, Urry95}). The
broad-band electromagnetic spectrum is composed of a synchrotron low-energy component that peaks (in a
$Log(\nu f(\nu))-Log(\nu)$ representation) between the far infrared and the X-ray band, followed by an
Inverse Compton component that has its maximum in the hard X-ray band or at higher energies, depending
on the location of the synchrotron peak, and extends into the $\gamma$-ray or even the TeV band. Those
Blazars where the synchrotron peak is located at low energy are usually called Low energy peaked
Blazars or LBL, while those where the synchrotron component reaches the X-ray band are called High
energy peaked Blazars or HBL [see Fig. \ref{all_bl_sed} and \cite{P95}]. LBL sources are the large
majority among Blazars [e.g., \cite{Pad03}] and are usually discovered in radio surveys, while HBL
objects are preferentially found in X-ray flux limited surveys since at these frequencies they are
hundreds, or even thousands, of times brighter than LBLs of similar radio power.

Radio quiet AGN are approximately one order of magnitude more abundant than Blazars and have been
shown to be the major constituent of the Cosmic X-ray Background (CXB) (\cite{Giacconi62, Rosati02,
Moretti03}) leading to the now consolidated picture in which the CXB is composed of radiation
generated by the accretion onto super-massive black holes integrated over cosmic time.

Despite the relatively low space density of Blazars, their strong emission across the entire
electromagnetic spectrum makes them potential candidates as significant contributors to extragalactic
Cosmic Backgrounds at frequencies where the accretion mechanism does not produce much radiation.
These extragalactic backgrounds would then be mostly composed of non-thermal radiation generated in
Synchrotron/Inverse Compton-type environments.

Recently, \cite{giocol04} showed that Blazars are by far the largest population of extragalactic
objects detected as foreground sources in CMB anisotropy maps and that their emission contaminates the
CMB angular power spectrum at a significant level. \cite{P93}, based on the detection of a small
sample of Blazars at $\gamma$-ray frequencies concluded that the Blazar population should produce a
large fraction of the high energy Cosmic Background.

In this paper we re-assess the Blazar contribution to the Cosmic energy in the microwave (CMB),
in the X-ray (CXB), in the Gamma-ray (CGB) and TeV (CTB) part of the electromagnetic spectrum.
Our calculations rely on a new estimation of the Blazar radio LogN-LogS, that we have assembled
combining several radio and multi-frequency surveys, on flux ratios in different energy bands and
on observed Blazar broad-band Spectral Energy Distributions (SED).

\section{The Blazar radio LogN-LogS}

In this section we use  several radio and multi-frequency surveys to build a new deep Blazar radio
LogN-LogS that updates and extends to lower fluxes the counts presented in \cite{giocol04}. Since the
surveys considered have been carried out at three different observing frequencies (1.4, 2.7 and 5
GHz), we convert all flux densities to a common frequency before proceeding. We take 5 GHz as the
reference frequency and (unless otherwise stated) we apply the flux conversions assuming a spectral
slope \ar = 0.25 ($f \propto \nu^{-\alpha_{\rm r}}$) which is approximately equal to the average value
in all our samples.

The derived Blazar counts are shown in Fig. \ref{logns} and can be described
by a broken power law with parameters defined in the following equation:


\begin{equation}
\displaystyle {N(>S)} =
 \cases{
 5.95~10^{-3}\times S^{-1.62} &  $S > 0.015~Jy$ \cr
 0.125\times S^{-0.9}         &  $S < 0.015~Jy$ \cr
 }
\label{eq.L1}
\end{equation}

The slope up to 15 mJy (dot-dashed line) is a good representation of the data, whereas the flattening
at fainter fluxes (dotted line) is necessary to avoid that the predicted Blazar space density exceeds
the total density of NVSS radio sources at a few mJy. The slope below the break is somewhat arbitrary
as only upper limits are available in this flux regime. We have chosen a value of 0.9 since this is
the average slope of the LogN-LogS of radio quiet AGN in the two flux decades below the break
(\cite{Rosati02, Moretti03}) and is consistent with all the available constraints.

In the following we discuss the details of all the surveys in order of decreasing radio flux limit.

\subsection{Two Jansky flat spectrum radio survey}

\begin{figure}[ht]
\centerline{
\includegraphics[width=6.5cm, angle=-90] {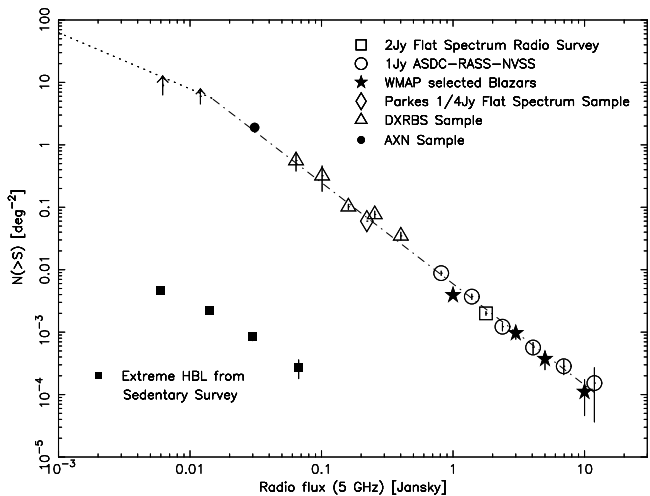}
}
\caption{The radio (5GHz) LogN-LogS of Blazars built combining several radio and multi-frequency
surveys. The different symbols represent the surveys used (see text for details). The filled square symbols in the lower part of the diagram represent the LogN-LogS of extreme HBL BL Lacs from the Sedentary Survey. Although these objects are a tiny minority ($\approx 0.1 \%$) of the overall Blazar population they play an important role in producing the Cosmic Background at X-ray and at very high energies ($\gamma$-ray/TeV).}

\label{logns}
\end{figure}

The two Jansky 2.7 GHz sample (\cite{Wall85}) is based on a complete radio flux limited survey of flat
spectrum ($\alpha_{\rm r} < 0.5$, $f \propto \nu^{-\alpha_{\rm r}}$ ) sources covering the entire sky
with the exclusion of the Galactic plane ($|b|> 10$). The sample includes 60 Blazars (\cite{diS94,
Urry95}) corresponding to a space density of 0.002 deg$^{-2}$. This value has been plotted as an open
square symbol in Fig. \ref{logns}.

\subsection{One Jansky ASDC-RASS-NVSS Blazar Sample}

The ASDC-RASS-NVSS 1Jy Blazar (1Jy-ARN) Survey (\cite{Giommi02}) is a radio flux limited ($f_r > 1 Jy
@1.4$GHz ) sample of Blazars selected by means of a cross correlation between the ROSAT All Sky Survey
(RASS) catalog of X-ray sources (\cite{Vog99}) and the subsample of NVSS survey (\cite{Con98}) sources
with flux density larger than 1 Jansky. The selection was carried out applying the same
multi-frequency technique used for the definition of the extreme HBL BL Lac sample in the Sedentary
survey (\cite{Gio99}). To avoid the complications due to high $N_H$ photoelectric absorption close to
the Galactic plane, only sources located at Galactic latitudes larger than 20 degrees were considered.
The accurate radio positions allowed us to obtain firm associations with all the optical counterparts.
The sample includes a total of 226 sources, 96\% of which were previously known objects, in  most
cases well documented in the literature as they are bright radio sources. The class composition of the
sample is reported in Table 1.

\begin{table}[h*]
\caption{One Jansky ASDC-RASS-NVSS sources}

\begin{tabular}{cl}
\hline
\multicolumn{1}{c}{Number} &
\multicolumn{1}{c}{Classification} \\
 \multicolumn{1}{r}{of sources} &
 \multicolumn{1}{r}{ } \\
\hline
132& FSRQs\\
25& BL Lac \\
36& SSRQs \\
21& radio galaxies\\
3& other galaxies\\
9& unidentified (3 flat, 6 steep) \\
\hline
\end{tabular}
\label{tab.ARN}
\end{table}

From past X-ray measurements we know that Blazars are characterized by a \fxfr ratio ranging from
$\approx 1.\times 10^{-13}$ \ergj to over $ 1.\times 10^{-9}$ \ergj (\cite{Pad02}) and therefore all
Blazars with flux larger than 1 Jy should be detectable in a survey with X-ray sensitivity of
$1.\times 10^{-13}$ \ergj. However, the X-ray limiting flux of the RASS strongly varies across the sky
depending on the effective exposure and on the amount of Galactic absorption ($N_H$) along the line of
sight.

\begin{figure}[ht]
\vbox{
\centerline{
\includegraphics[width=6.5cm, angle=-90] {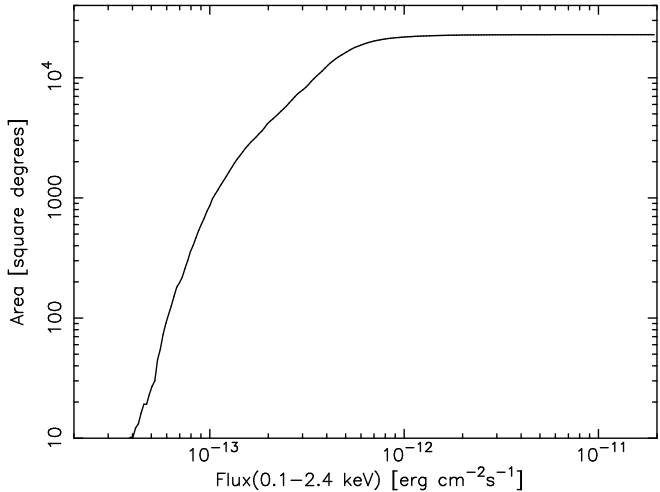}
}}
\caption{The high Galactic Latitudes ($|b| > 20$) sky coverage of the RASS survey in the region
of overlap with the NVSS catalog (Dec $>$ -40 degrees).}
\label{rass_skycov}
\end{figure}

Figure \ref{rass_skycov} shows the high Galactic Latitude ($|b| > 20$)  sky coverage of the RASS in the
region of overlap with the NVSS survey. The curve was calculated taking into account of the ROSAT effective exposure in sky bins of 1 square degrees in size and converted to sensitivity assuming a power law spectrum with energy index  $\alpha=1.0$ and setting the amount of $N_H$ equal to the Galactic value in the direction of each sky bin.

The X-ray sensitivity ranges from somewhat below $1\times 10^{-13}$ \ergs, where
about 1000 square degrees of sky are covered, to $\approx 2.\times 10^{-12}$ \ergs,
where the maximum area of 23,000 square degrees is reached.

Assuming that the three unidentified flat spectrum sources are Blazars,
the number of objects of this type (FSRQ+BL Lacs+unidentified sources with \ar $<$ 0.5) in
the 1Jy-ARN survey is 160.

This sample has been used to estimate the Blazar space density above 1Jy taking into account that the
sky coverage of Fig. \ref{rass_skycov} implies that faint sources with flux around e.g. $1. \times
10^{-13}$ \ergs could be detected only in  $\approx 1,000$  square degrees of sky, whereas brighter
X-ray sources with flux of e.g. $1. \times 10^{-12}$ \ergs could be detected over a much larger
portion of the sky ($\approx 20,000$ square degrees). The counts have been converted to 5GHz and then
plotted as open circles in Fig. \ref{logns}.

\subsection{WMAP microwave selected Blazars}

The Wilkinson Microwave Anisotropy Probe (WMAP) is a space observatory dedicated to the accurate
investigation of primordial fluctuations in the Cosmic Microwave Background (\cite{bennett03a}). A
catalog of 208 bright foreground sources detected in one or more of the five microwave WMAP channels
during the first year all sky survey has been published by \cite{bennett03}. With very few exceptions
all entries are well known bright sources at cm wavelengths and in most cases observed at several
radio frequencies. \cite{giocol04} have recently shown that the large majority of these objects are
Blazars.

To determine how WMAP selected Blazars contribute to the 5GHz counts we have considered all
high Galactic latitude ($|b| > 20 $ deg) detections with flat radio spectrum (\ar $<$ 0.5,
$f \propto \nu^{-\alpha_{\rm r}}$) and with SED typical of Blazars.
Table \ref{tab.wmap} gives the detailed statistics:

\begin{table}[h]
\caption{WMAP bright foreground source catalog}

\begin{tabular}{cl}
\hline
\multicolumn{1}{c}{Number} &
\multicolumn{1}{c}{Classification} \\
 \multicolumn{1}{r}{of sources} &
 \multicolumn{1}{r}{ } \\
\hline
141& FSRQs\\
23& BL Lac \\
13& radio galaxies\\
5& SSRQs \\
2& starburst galaxies\\
2& planetary nebulae \\
17& unidentified \\
5& no radio counterpart \\
& (probably spurious) \\
\hline
\end{tabular}
\label{tab.wmap}
\end{table}

The space density of WMAP detected sources with radio flux above 1, 3, 5 and 10 Jansky at 5GHz
has been plotted in Fig. \ref{logns} as filled star symbols.

With the exception of the point at 1 Jansky, which is most likely an underestimation due to
incompleteness of the WMAP catalog at this flux limit (\cite{bennett03}), the agreement with other
radio surveys at cm wavelengths is very good implying that the Blazar selected by WMAP and at radio
frequencies belong to the same population of objects.

\subsection{The Parkes quarter-Jansky Flat Spectrum sample}

The Parkes quarter-Jansky Flat Spectrum sample (\cite{wall04}) is a 100\% identified radio flux
limited survey at a frequency of 2.7 GHz.

A total of 328 FSRQ and 43 BL Lacs have been detected with flux density higher than 0.25 Jy in an area
of 8785 square degrees (sample 1 of \cite{wall04}). Considering that this survey only accepts sources
with spectral index flatter than \ar $< 0.4$ ($f \propto \nu^{-\alpha_{\rm r}}$) we rescale the sample
density by a factor 1/0.75 which is the ratio of QSOs and BL Lacs with \ar $< 0.7$ to those with \ar
$< 0.4$ in the Parkes survey known as PKSCAT90 with flux density larger than 0.25 Jy. A similar ratio
is present in the 1Jy-ARN survey.

The space density of Blazar in this survey is therefore 0.06 objects per square
degree. We convert 2.7GHz fluxes to 5GHz  and plot the density
as an open diamond in Fig. \ref{logns}.

We note that 84 additional sources are classified by \cite{wall04} as flat radio spectrum Galaxies (57
of which without redshift); as some of these may well be BL Lacs the Blazar content of this sample
will probably grow in the future.

\subsection{The DXRBS Blazar Survey}

The Deep {X}-Ray Radio Blazar Survey (DXRBS) is a radio flux limited sample based on a double
selection technique at radio and X-ray frequencies and uses optical data to refine the sample.  DXRBS
searches for Blazar candidates among serendipitous X-ray sources of the ROSAT PSPC pointed
observations listed in the WGA catalog (\cite{WGA95}), restricting the sample to objects with radio
spectral index flatter than \ar = 0.5 and with broad-band spectral indices \aro and \aox in the region
occupied by Blazars. Details of the selection method and of the optical identification of the
candidates are described in \cite{Per98, L01,Pad05}.

Although the radio limit of the survey is 50 mJy at 5GHz, the X-ray sensitivity of the ROSAT PSPC
pointings  is not deep enough to ensure completeness at all radio fluxes, as some Blazars with flux
lower than 100 mJy are expected to have an X-ray flux below the ROSAT WGA limit of $ \approx 2 \times
10^{-14}$ \ergs. We have therefore estimated the  Blazar space density taking into account the latest
LogN-LogS results of  \cite{Pad05}  for both FSRQs and BL Lacs and correcting the points at 50 and 100
mJy for the fraction of lost objects as predicted from the distribution of Fig. \ref{fxfr} .

The final Blazar space density from this survey is plotted in Fig. \ref{logns} as open triangles.

\subsection{The ASDC-XMM-Newton-NVSS (AXN) Sample}

In order to push the radio limit to fluxes significantly below 50 mJy at 5GHz using the same
multi-frequency selection method of the DXRBS survey it is necessary to reach X-ray sensitivities
proportionately deeper than that of the ROSAT-PSPC. For this purpose, we have searched for
serendipitous NVSS radio sources in XMM-Newton EPIC-pn (\cite{Strueder01}) X-ray images which provide
approximately one order of magnitude better sensitivity than ROSAT.

We have used the data in the XMM-Newton public archive available at the ASI Science Data Center (ASDC) in December 2004.
At that date a standard processing, which removes periods of high background resulting from solar flares and detects serendipitous sources, had been run at ASDC on 1220 EPIC-pn fields, 847 of which
at high galactic latitude ($|b| > 20$).

Of these EPIC-pn observations we have considered the subsample of 188 non-overlapping fields that satisfy the conditions listed below that are necessary to ensure that the sample of serendipitous sources is suitable for our purposes:

\begin{enumerate}
\item the EPIC-pn instrument was used in full window mode;
\item the pointing of the XMM observation was at declination north of -40 degreed to ensure overlap
      with NVSS survey;
\item no bright or extended source is present in the field of view;
\item whenever more than one exposure of the same field was available in the archive, the deeper one was taken.
\end{enumerate}

In addition, we have excluded a 5 arc-minutes circular area around the target of each field. The total
area covered by the 188 fields is 26.3 square degrees. A careful inspection of all the
X-ray/radio/optical associations with broad-band spectral indices \aro and \aox in the region occupied
by Blazars  (\cite{Gio99}) led to the results summarized in Table \ref{tab.xmm}.

\begin{table}[h]
\caption{NVSS radio sources detected in XMM-Newton EPIC-pn X-ray images}
\begin{center}
\begin{tabular}{cccc}
\hline
\multicolumn{1}{c}{Number} &
\multicolumn{1}{c}{Space density} &
\multicolumn{1}{r}{Flux limit} &
\multicolumn{1}{r}{Flux limit} \\
 \multicolumn{1}{c}{of sources} &
 \multicolumn{1}{c}{sources deg$^{-1}$} &
 \multicolumn{1}{c}{1.4 GHz} &
 \multicolumn{1}{c}{5 GHz $^*$} \\
\hline
50& $1.9\pm 0.3$ & 50&31 \\
107& $>4.0$ &20&12  \\
149& $>5.6$ &10&6  \\
\hline
\end{tabular}
\label{tab.xmm}
\end{center}
$^*$ We assume a spectral slope \ar $=0.4$

\end{table}

  For this first (conservative) estimation we assume that all XMM fields are sufficiently deep to detect all Blazars above 50 mJy. Because of the spread in exposure times and of the reduction in sensitivity at large off-axis angles due to vignetting effects and PSF degradation,  some objects (of the order of 20-30\%) might have been missed and the number of Blazars found is only a lower limit. On the other hand, from the
1Jy-ARN and other surveys, we expect that a similar percentage of candidates be associated to steep
spectrum objects, so that the two effects roughly cancel out. We will discuss the impact of this
reduction in sensitivity in a future paper (\cite{Cavazzuti05}) where the sky coverage of the sample
will be fully taken into account.

The space density at 50 mJy at 1.4GHz (31 mJy at 5.GHz) has been plotted in Fig. \ref{logns} as a filled
circle.
The points at 12 and 6 mJy are drawn as lower limits because in this case a significant fraction (e.g. $\gsim $50\%) of these faint Blazars (especially at 6 mJy) is expected to be below the sensitivity limit of our XMM-Newton X-ray images.
In fact, the \fxfr distribution of known Blazars (see Fig. \ref{fxfr}) implies that the expected
soft X-ray flux of a 10 mJy Blazar is below $2\times 10^{-15}$ \ergs in about 30\% of the cases.
These X-ray sources could not be detected in most of our EPIC images, especially at large off-axis angles where most of the area is located.

\subsection{The Sedentary Survey of extreme High Energy Peaked BL Lacs}

We conclude this section summarizing some of the results from the multi-frequency `Sedentary Survey'
(\cite{Gio99, paperIII, Gio05}) which is a deep ($f_r \ge 3.5 $ mJy at 1.4 GHz), 100\% identified
radio flux limited sample of 150 extreme HBL objects characterized by a \fxfr ratio higher than
$3\times 10^{-10}$ \ergj. This survey does not have a direct impact on the full Blazar LogN-LogS of
Fig. \ref{logns} since such extreme HBL objects only represent a tiny minority of the full Blazar
population at radio frequencies. However, the very high \fxfr flux ratio, which in these sources
ranges from five hundred to over five thousand times that of typical Blazars, makes these rare radio
emitters a numerically important population of sources at X-ray and TeV frequencies and therefore
makes them potentially significant contributors to the Cosmic Background radiation at these
frequencies. The space density of the objects at fluxes $f_r \ge 6  $ mJy (5 GHz) in this survey is
reported in Fig. \ref{logns} as filled squares.

\section {From radio frequencies to other spectral bands}

Once the LogN-LogS of a population of sources is known in a given energy band,
it is possible to estimate their emission in other parts of the electromagnetic spectrum
provided that the flux ratio in the two bands, or even better, the overall energy distribution,
is known.

In this section we deal with flux ratios and SEDs of Blazars that will be used
later in the paper to estimate the contribution to Cosmic Backgrounds at frequencies higher than the radio ones.

The collection of SED of a large sample of Blazars built using \sax and multi-frequency data presented
by \cite{gio02c} clearly shows that a wide variety of broad-band spectral forms exist. This large
spectral diversity, however, can be reproduced, at least in first approximation, by Synchrotron Self
Compton (SSC) emission models as those shown in Fig. \ref{all_bl_sed}. Sometimes, to explain the
$\gamma$-ray part of the distribution, it is necessary to add other components to this simple picture.
We will deal with this possibility later when we consider the contribution of Blazars to the
$\gamma$-ray background.

Although a continuous range of synchrotron peak energies probably exists, it is useful to divide the population
of Blazars into Low energy peaked Blazars (or LBL) whose X-ray emission is due to Inverse Compton radiation,
and in High energy peaked Blazars (HBL) characterized by a much higher X-ray flux (for the same radio flux)
since their synchrotron component extends into the X-ray band.
This simple separation in two broad categories also reflects the early selection methods
since LBL objects have mostly been discovered in radio surveys, while HBL objects have been typically
found in X-ray flux limited samples.

\begin{figure}[ht]
\vbox{
\centerline{
\includegraphics[width=9.cm, angle=0] {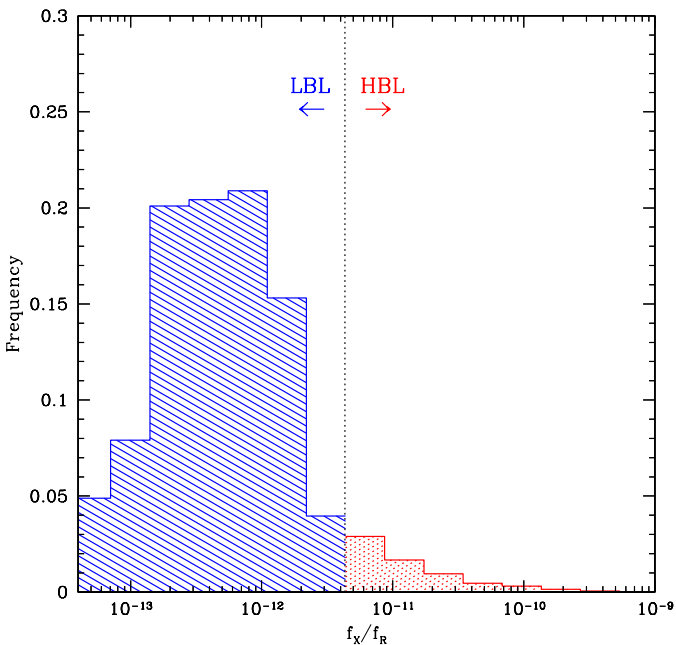}
}}
\caption{The \fxfr distribution of Blazars estimated from the 1Jy-ANR and from a sample of about 2000 HBL Blazar candidates (see text for details). }
\label{fxfr}
\end{figure}

The large difference in X-ray emission in LBL and HBL objects is graphically reflected in the distribution
of \fxfr flux ratios shown in Fig. \ref{fxfr} which spans about 4 orders of magnitudes.

The plot shown in Fig. \ref{fxfr} was built using two radio flux limited samples. At \fxfr values
lower than $5\times 10^{-12}$ \ergj  (corresponding to LBL objects) we have used the 1Jy-ARN sample
corrected for the RASS sky coverage. As the percentage of Blazars sharply drops at \fxfr values larger
than $5\times 10^{-12}$ \ergj the 1Jy-ARN sample rapidly becomes statistically inadequate and lager
samples are clearly needed. We have therefore built a radio flux limited sample including nearly 2000
Blazar candidates with \fxfr $> 5\times 10^{-12}$ \ergj applying the same technique used for the
Sedentary survey which is about 85\% efficient in selecting HBL Blazars with \fxfr $> 3\times
10^{-10}$ \ergj (\cite{Gio05}).

To that purpose, we have cross-correlated the positions of the X-ray sources in the ROSAT All Sky
Survey (\cite{Vog99}) with the radio sources of the NVSS catalog (\cite{Con98}) and we have estimated
optical (J or F) magnitudes from the GSC2 Guide Star Catalog (\cite{McLean00}). We have then
calculated the broad-band \aox and \aro spectral slopes and we have only accepted objects in the area
of the \aox - \aro plane occupied by Blazars (\cite{Gio99}). Whenever no optical counterpart was found
within the radio positional uncertainty we have assumed a lower limit of Jmag = 19.5, which is the
limit of the GSC2 catalog.

To estimate the distribution at  \fxfr  $> 5\times 10^{-12}$ \ergj the radio flux limit of the sample
must correspond to an X-ray flux at which the ROSAT survey covers a sky area large enough to
allow the detection of at least a few objects.
We have chosen $f_r=$25 mJy, corresponding to $f_X > 1.25 \times 10^{-13}$ \ergs where the RASS
covers about 2000 square degrees of sky. We have then derived the tail of the \fxfr  distribution for \fxfr $< 5\times 10^{-12}$  \ergj taking into account the RASS sky coverage shown in Fig. \ref{rass_skycov}.

We have checked the reliability of this selection method using the subsample of 514 objects for which
Sloan Digital Sky Survey (SDSS) [e.g., \cite{York00, Stoughton02}] spectral data (\cite{turr05}) are
available and we have found that over 80\% of the 319 candidates with \fxfr $> 5\times 10^{-12}$ \ergj
are indeed spectroscopically confirmed Blazars.

To properly normalize this distribution we scaled it so that the fraction of Blazars with
\fxfr  $> 3\times 10^{-10}$ \ergj (the \fxfr limit of the Sedentary survey) is equal to the density
ratio between the full Blazar population and that of the extreme HBL of the sedentary survey
(dashed line and filled squares in Fig. \ref{logns}).

The combined \fxfr distribution is shown in Fig. \ref{fxfr} where we see that
the large majority (94\%) of Blazars are of the LBL type (defined here as objects with
\fxfr $< 5\times 10^{-12}$ \ergj).

\begin{figure}[ht]
\vbox{
\centerline{
\includegraphics[width=6.5truecm, angle=-90] {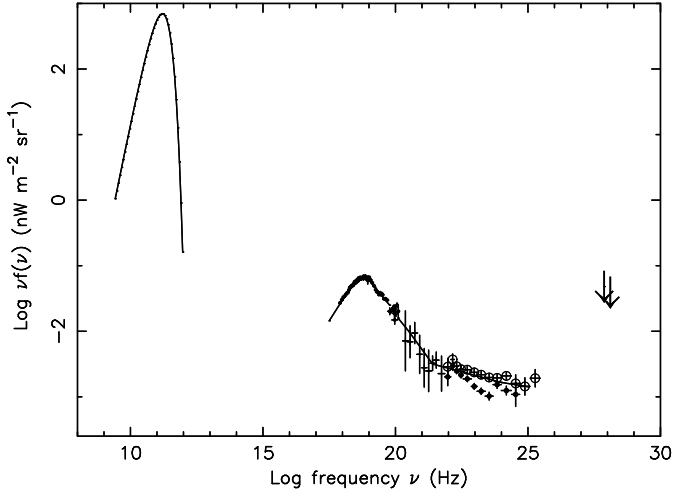}
}}

\caption{The extragalactic Cosmic Background Energy distribution at microwave, X-ray
and $\gamma$-ray energies.
}
\label{CBGs}
\end{figure}

\section {The contribution of Blazar to Cosmic Backgrounds}

Figure \ref{CBGs} shows the spectral energy distribution of the Cosmic Extragalactic Background radiation in
the microwave, X-ray and $\gamma$-ray band, that is where we expect that the Blazar collective emission gives a
significant contribution.

The Cosmic Microwave Background (CMB), taken from the very accurate measurements of the COBE satellite
is represented by a black-body spectrum with temperature of 2.725 $^o$K (\cite{Mather99}); the X-ray
Background (CXB) is taken from HEAO-1 measurements (\cite{Marshall80, Gruber99}) and has been scaled
to match the more recent \sax, ASCA and XMM-Newton results in the softer 2-10 keV band
(\cite{Vecchi99, Lumb02, Kushino02}). The Gamma Ray Background is derived from the COMPTEL data in the
range $0.8 - 30$ MeV (\cite{Kappadath98}), and from EGRET data in the range $30$ MeV - $50$ GeV. We
recall that the estimate of the extragalactic  $\gamma$-ray background emission depends on the
Galactic diffuse emission model which itself is not yet firmly established. In Fig. \ref{CBGs} we
report the results of two different analysis of the EGRET data: open circles from \cite{Sreekumar98}
and filled circles from \cite{Strong04}, which uses an improved model of the Galactic diffuse
continuum gamma-rays. As for the TeV diffuse background, we report the upper limits in the $20$ -
$100$ TeV region derived from the HEGRA air shower data analysis (\cite{hegra01}). Note that the HEGRA
measurements are sensitive both to non-isotropic (galactic) and isotropic (extragalactic) component.
In the 1 TeV -- 1 PeV energy range, other experiments give only upper limits and there is no clear
observation of a diffuse photon signal yet.

In the following we estimate the Blazar contribution to the above described Cosmic Backgrounds
basing our calculation on the radio LogN-LogS built in Sect. 2 and on flux ratios in different
bands or on the observed Spectral Energy Distributions of $\gamma$-ray detected sources.

\subsection{The Cosmic Microwave Background}

The contribution of Blazars to the Cosmic Microwave Background (CMB) has been estimated in the past
from different viewpoints (see, e.g., \cite{toff98, giocol04}).

Here we use the Blazar radio LogN-LogS of Fig. \ref{logns}  to update the results presented in
\cite{giocol04}.

From Fig. \ref{logns} we see that the Blazar LogN-LogS can be represented by a broken power law model
 with  alpha=1.62 (integral slope) up to a break point where the slope flattens significantly.
The precise position of the break and the amount of flattening cannot be estimated with our data.
However, a break must occur somewhere  around 10-15 mJy otherwise the number of Blazars would exceed
the total number of radio sources at about 2-3 mJy. In the following we use the somewhat conservative
values of 15 mJy for the break and 0.9 for the slope below the break. Both values are consistent with
the lower limits on the Blazar counts at 12 and 6 mJy (see Sect. 2)

The integrated background intensity due to Blazars therefore can be expressed as
\begin{equation}
 \displaystyle {I_{Blazars}=\int_{0.1mJy}^{1Jy} S~{dN \over dS}~dS } ~,
 \label{eq.IBLazars}
\end{equation}
where dN/dS is the differential of eq. (\ref{eq.L1}). The minimum integration flux of 0.1 mJy for
$S_{min}$ is likely to be conservative since Blazars with radio flux near or below 1 mJy are already
included in the {\it Einstein} Medium Sensitivity Survey BL Lac sample (\cite{Rec00}). The integrated
intensity $I_{Blazars}$ is then converted from 5GHz to microwave frequencies convolving the flux value
with the observed distribution of spectral slopes between 5GHz to microwave frequencies as estimated
from the 1Jy-ARN sample  (see \cite{giocol04} for details).

\begin{figure}[ht]
\vbox{
\centerline{
\includegraphics[width=7.0cm, angle=-90] {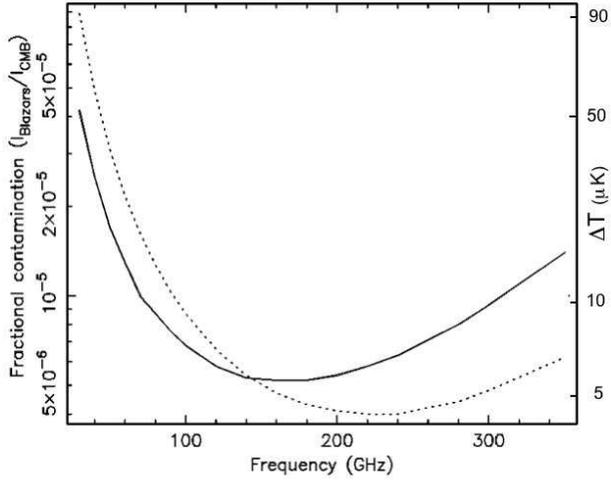}
}}

\caption{The Blazar fractional contamination of the CMB (solid line) and the apparent
increase in CMB temperature due to Blazars (dotted line) as a function of frequency}
\label{contamination}
\end{figure}

Figure \ref{contamination} plots the fractional contamination, defined as $I_{Blazars}/I_{CMB}$,
that is the ratio between the Blazar integrated emission and the CMB intensity (solid line), and
the equivalent apparent temperature increase of the CMB (dotted line), as a function of frequency.

\begin{figure}[ht]
\vbox{
\centerline{
\includegraphics[width=8.0cm, angle=0] {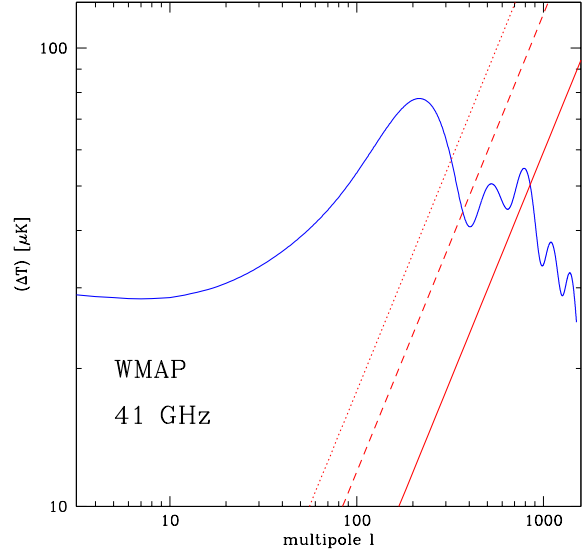}
}}

\caption{The contribution of Blazars to the CMB fluctuation spectrum in the WMAP Q channel at 41 GHz
as evaluated from the LogN-LogS given in Fig. \ref{logns} (solid line). We also show the angular power spectrum
for the Blazar population by adding an estimate of the possible contribution of radio sources with
steep-spectrum at low radio frequencies which flatten at higher frequencies (dashed line). The dotted
line includes also the effect of spectral and flux variability (see text for details). Although this
additional contamination may be substantial a precise estimation can only be done through simultaneous
high resolution observations at the same frequency. A typical CMB power spectrum evaluated in a
$\Lambda$CDM cosmology with $\Omega_m=0.3, \Omega_{\Lambda}=0.65, \Omega_b=0.05$ which best fits the
available data is shown for comparison. } \label{cmb41}
\end{figure}

\begin{figure}[ht]
\vbox{
\centerline{
\includegraphics[width=8.0cm, angle=0] {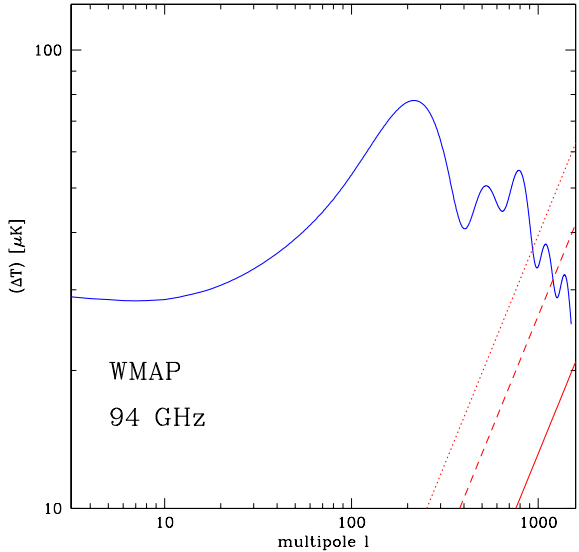}
}}

\caption{The contribution of Blazars to the CMB fluctuation spectrum as in Fig. \ref{cmb41} for
the WMAP 94 GHz channel.
}
\label{cmb94}
\end{figure}

The contribution of Blazars to the temperature anisotropy spectrum of the CMB is calculated as \\
$(\Delta T)_{Blazar}= [(2 \pi)^{-1} C_{\ell} \ell (\ell +1)]^{1/2}$, where
 \be
 C_{\ell, {\rm Blazar}} = \int_{S_{\rm min}}^{S_{\rm max}} dS~{dN \over dS}~S^2 ~,
 \label{Eq.cl}
 \ee
assuming that the Blazars are spatially distributed like a Poissonian random sample (\cite{Tegmark96,
Scott99, giocol04}). The quantity at the right hand side in eq. (\ref{Eq.cl}) is the usual Poisson
shot-noise term (\cite{Peebles80, Tegmark96}). We neglect here the clustering term, $\omega(I)^2$,
which adds to the Poissonian term since there is not yet a clear estimate of the Blazar clustering. We
note however, that the inclusion of clustering can significantly increase the amount of contamination,
especially at large angular scales (\cite{Gonzales05}).
\\
For the Blazar population described by the LogN-LogS given in fig. \ref{logns}, we found $C_{\ell, {\rm
Blazar}} \approx 54.4 ~ {\rm Jy}^2 {\rm sr}^{-1}$ at 41 GHz and $C_{\ell, {\rm Blazar}} \approx 51.6~{\rm Jy}^2 {\rm sr}^{-1}$ at 94 GHz. Our results can be translated into temperature units using the conversion between the isotropic black-body (Planckian)
brightness $B_0(\nu)$ and the CMB temperature $T_0$ which can be written as
\begin{eqnarray}
 {\partial B_0 \over \partial T_0} &=& {k \over 2} \bigg( {k T_0 \over h c}
 \bigg)^2 \times \bigg[ {x^2 \over sinh(x/2)}\bigg]^2 \nonumber\\
 &=& \bigg( {24.8 Jy ~sr^{-1} \over \mu {\rm K}} \bigg)
 \times \bigg[ {x^2 \over sinh(x/2)}\bigg]^2 ~.
 \label{Eq.conv}
 \end{eqnarray}
 Here $x \equiv h \nu/k T_0 = \nu / 56.84$ GHz is the a-dimensional frequency given in
terms of the Planck constant $h$, of the speed of light $c$ and of the Boltzmann constant $k$.
We found a value $C_{\ell, {\rm Blazar}} \approx 2.22\cdot 10^{-2} ~\mu {\rm K}^2 sr$ at 41 GHz and
$C_{\ell, {\rm Blazar}} \approx 1.09 \cdot 10^{-3} ~\mu {\rm K}^2 sr$ at 94 GHz. We show in Figs. \ref{cmb41}
and \ref{cmb94} the quantity $\Delta T_{\rm Blazar}$ and we compare it to the CMB fluctuation power spectrum which
best fits the available data.
The previous value is shown as solid lines and has to be considered as a definite lower limit for $C_{\ell,
{\rm Blazar}}$ since it neglects the contribution of steep-spectrum sources at low frequencies which
flatten at these frequencies (41 and 94 GHz) and the effect of flux variability. The dashed lines in Figs. \ref{cmb41}
and \ref{cmb94} show the Blazar power
spectrum by adding the contribution of radio sources with steep-spectrum at low radio frequencies
which flatten at higher frequencies. The effect of spectral and flux
variability also allows for an increase of the fluctuation level $C_{\ell, {\rm Blazar}}$
because many Blazars below the sensitivity threshold of CMB experiments can be detected
during flares. The Blazar flux variability at millimeter wavelengths may be very substantial (higher
than factors 3-10 on time scales of weeks to years seen at cm wavelengths) and could definitely
increase the level of contamination of CMB maps when these are built over long integration periods.
The dotted lines in Figs. \ref{cmb41} and \ref{cmb94} show the Blazar power spectrum when also this last contribution is taken into account.\\

Our numerical estimates are subject to some uncertainties which depend on the parameters of the
LogN-LogS that are not well constrained.  For instance if instead of using the conservative values of 15
mJy and 0.9 for the break flux and slope below the break we used the less conservative values of 10
mJy and $\alpha= 1.0$ we would have an increase of $ \approx $ 50\% for  $I_{Blazars}$ and of
$\approx $4\% for $C_{\ell,{\rm Blazar}}$ .

The contamination level shown in  Figs. \ref{cmb41} and \ref{cmb94} is that expected in case no
Blazars are removed from the CMB data. The observed WMAP angular power spectrum reported by Hinshaw
(2003) has been obtained using a procedure in which a pure primordial signal has been initially
assumed, and then subtracting the foreground sources that could be recognized as such in the WMAP
data, thus removing only the bright tail of the Blazar LogN-LogS distribution. The procedure adopted
to derive the CMB power spectrum should take into account the full point-like source contribution
implied by our LogN-LogS. Such an approach would both influence the shape of the expected power
spectrum and increase the statistical uncertainties of the WMAP data points, especially at high
multipoles, where the Blazar contribution is larger.

We stress that the previous calculations are performed neglecting the clustering term $\omega(I)^2$
and thus they must be considered again as a lower limit to the realistic angular power spectrum
contributed by the Blazars. The effects of clustering of FSRQ on the CMB fluctuation spectrum has been
partially estimated by some authors: for instance \cite{Gonzales05}  calculated from simulations that
the clustering of extragalactic radio sources is rather small at $\ell = 200$ for the Planck
frequencies, while the contribution of the clustering term to the confusion noise is likely to be the
dominant one. Scott and White (1999) estimated instead that the clustering of SCUBA sources to the
expected Planck CMB spectrum provides an increase of a factor 5-10 with respect to the Poissonian term
for the point-like source contribution at $\ell =200$. It is clear that the expectations for the
clustering effect strongly depend on the adopted model for the source counts and on their clustering
model. Based on the correlation function of \cite{Loan97} (adopted by Toffolatti) and on that for the
SCUBA sources adopted by \cite{Scott99} and using our Blazar LogN-LogS, we expect that the
contamination of the first peak of the fluctuation spectrum (at the WMAP 41 GHz channel) is at a level
comprised in the range 20-25 \%. This estimate does not include possible variability effects and
additional core-dominated radio sources

\subsection{The Soft X-ray Background}

\begin{figure}[ht]
\vbox{
\centerline{
\includegraphics[width=6.5cm, angle=-90] {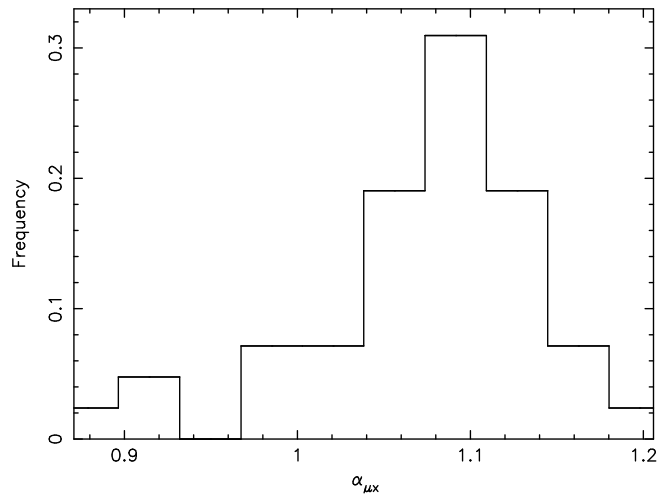}
}}
\caption{The distribution of spectral slopes between the microwave (94 GHz) and X-ray
(1keV) band measured from the sample of 42 Blazars included in the WMAP catalog of bright
microwave sources detected in the 94 GHz channel and for which X-ray measurements
are available.
}
\label{a94x}
\end{figure}

We estimate the contribution of Blazars to the CXB at 1 keV using two methods:
i) converting the integrated radio flux calculated with eq. (\ref{eq.IBLazars})
into X-ray flux using the observed distribution of \fxfr flux ratios of Fig. \ref{fxfr}; and
ii) converting the integrated Blazar contribution to the CMB (at 94 GHz) to X-ray flux using
the distribution of \amx, the microwave (94 GHz) to X-ray (1keV) spectral slope, defined as
\begin{equation}
 \displaystyle {\alpha_{\mu x} = -{ Log(f_{1keV}/f_{94GHz})\over{Log(\nu_{1keV}/\nu_{94GHz})}}
 = -{Log(f_{1keV}/f_{94GHz})\over{6.41}} }
 \label{eq.alphamux}
\end{equation}
and estimated from the subsample of Blazars detected by WMAP in the 94 GHz channel and
for which an X-ray measurement is available (see Fig. \ref{a94x}).

The first method gives a total Blazar contribution to the X-ray background of $2.7\times 10^{-12}$
\ergssd (about 70\% of which due to HBL sources with \fxfr  $> 5 \times 10^{-12}$ \ergj) in the ROSAT
0.1-2.4 keV energy band. Assuming an average Blazar X-ray energy spectral index of \ax = 0.7 (which is
a compromise between a flatter spectrum that is typical of LBL sources and an \ax steeper than 1 for
HBL objects) this flux converts to $2.6\times 10^{-12}$ \ergssd in the 2-10 keV band or 11\% of the
CXB which is estimated to be $2.3\times 10^{-11}$ \ergssd (\cite{perri00}).

The distribution of microwave to X-ray spectral slope  of Fig. \ref{a94x} has an
average value of $\langle$ \amx $\rangle$ =  1.07 and a standard deviation of 0.08
corresponding only to about a factor 3 in flux. This distribution is much narrower than
that between the radio and X-ray band (Fig. \ref{fxfr}, \fxfr) and the dispersion
is comparable to the variability that Blazars show at radio and especially at X-ray frequencies.

Since the 94 GHz and X-ray measurements are not simultaneous most of the dispersion in the
distribution shown in Fig. \ref{a94x} is probably due to variability both in the microwave and X-ray
band. The intrinsic dispersion is therefore likely to be much smaller. For this reason in converting
from the background level at 94 GHz estimated in the previous section to 1 keV we use the average \amx
value and assume no dispersion. We must note however that, just like the 1Jy radio surveys, the WMAP
\amx distribution hardly includes any HBL sources, which make only a few percent of the population.\\
From eq. (\ref{eq.alphamux}) we get
\begin{equation}
 \displaystyle {f_{1keV}=1.4\times 10^{-7}~f_{94GHz} } ~.
 \label{eq.fxf94}
\end{equation}
Since the Blazar integrated emission at 94 GHz is $7.2\times10^{-6}$CMB$_{94GHz}$ or 0.64 Jy/deg$^2$
and the Cosmic X-ray Background is $2.3\times 10^{-11}$ \ergs (\cite{perri00}) in the 2-10 keV band,
equivalent to 2.31$\mu$Jy/deg$^2$ at 1 keV, eq. (\ref{eq.fxf94}) for $f_{94GHz}$=0.64 Jy/deg$^2$ gives
f$_{1keV}$ = 0.09 $\mu$Jy/deg$^2$ or 3.9\% of the CXB.

Taking into account that the distribution of Fig. \ref{a94x} only includes LBL objects
and that HBL sources make two thirds the total contribution to the CXB (see above), the percentage
scales to about 12\% which is very close to the 11\% obtained with the previous method.

Both result are in good agreement with the independent estimate of \cite{Galbiati04}, based on the
XMM-Newton Bright Serendipitous Survey,  who conclude that the radio loud AGN content of the CXB is
13\%.

\subsection{Hard X-ray -- soft $\gamma$-ray Background}

The number of sources detected at energies higher than soft X-rays is still rather low
and building reliable distributions of flux ratios between radio or microwaves
and the Hard X-ray/$\gamma$-ray fluxes similar to those of Fig. \ref{fxfr} and \ref{a94x}
is not currently possible.
In order to estimate the Blazar contribution to high energy Cosmic Backgrounds (E$~>~$100 keV)
we have therefore followed a different approach. We have extrapolated the predicted
Blazar integrated intensity at microwave frequencies (eq. \ref{eq.IBLazars}) to the hard X-ray and
soft $\gamma$-ray band using a set of hypothetical Synchrotron Self-Compton Spectral Energy Distributions.

\begin{figure}[ht]
\vbox{
\centerline{
\includegraphics[width=7.0cm, angle=-90] {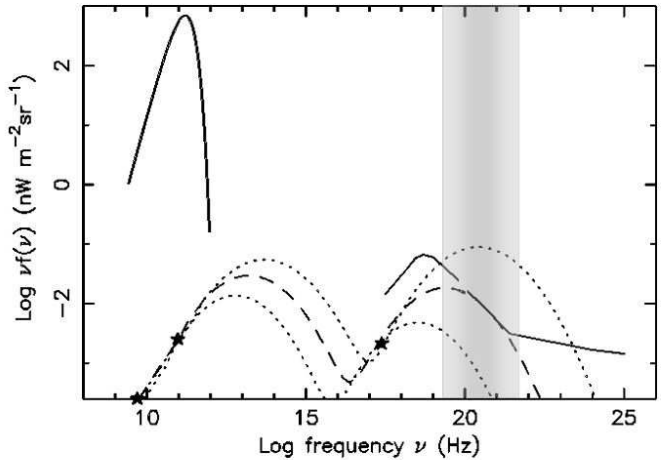}
}}
\caption{The possible contribution of LBL Blazars to the Hard X-ray/soft $\gamma$-ray
Background (shaded area). The three SSC curves corresponds to different $\nu_{peak}$ values
(log $\nu_{peak}= 12.8, 13.5 $and$ 13.8$) and are constrained to go through the
three star symbols representing 1) the total contribution of Blazars at 94 GHz
($8\times 10^{-6}$ of the CMB intensity), 2) the average 5 GHz to 94 GHz slope (\arm = 0.2)
and 3) the average spectral slope between 94 GHz and 1 keV ($\langle $ \amx  $\rangle = 1.07$).
}
\label{hardx_back}
\end{figure}

Figure \ref{hardx_back} shows the CMB, CXB and CGB together with three predicted
SED from a simple homogeneous SSC models whose parameters are constrained to
1) be consistent with the expected integrated flux at 94 GHz, 2) have the \amx slope equal
to the mean value of the WMAP Blazars (\amx = 1.07) and 3) possess a radio spectral slope
equal to the average value in the WMAP sample. The three curves, forced to pass through the
three star symbols graphically representing the three constraints listed above,
are characterized by synchrotron peak frequencies of \nupeak = $10^{12.8}$, $10^{13.5}$ and $10^{13.8}$ Hz.
From Fig. \ref{hardx_back} we see that a high value of \nupeak over-predicts by
a large factor the observed Hard-X-ray to soft $\gamma$-ray ($\approx 10^{20}-2\times10^{22}$ Hz or $\approx$ 500 keV-10 MeV)
Cosmic Background, whereas a too low value of \nupeak predicts a negligible contribution.
The case \nupeak = $10^{13.5}$ Hz predicts 100\% of the Cosmic Background. Since
the Log(\nupeak) values of Blazars in the 1Jy-ARN survey and WMAP catalog peak near
13.5 and range from 12.8-13.7 within one sigma from the mean value,
we argue that the data presently available indicate that Blazars may be responsible for a large fraction,
possibly 100\%, of the Hard-Xray/Soft $\gamma$-ray Cosmic Background.

\subsection{$\gamma$-ray Background}

The SSC distributions of Fig. \ref{hardx_back} predict a negligible Blazar contribution to the
extragalactic $\gamma$-ray Background above $100~$MeV. Nevertheless, it is well known that Blazars are
the large majority of the extragalactic $\gamma$-ray (E $>$ 100 MeV) identified sources detected by
the EGRET experiment (\cite{Hartman99}) aboard the Compton Gamma-ray Observatory, and therefore are
likely to contribute in a significant way to the $\gamma$-ray Background. Indeed \cite{P93}, on the
basis of a small number of sources detected by  EGRET, concluded that Blazars should make a large
fraction, if not the totality, of the extragalactic $\gamma$-ray background. However, these early
calculations relied upon a small database and had to assume no strong variability, a characteristics
that has been later demonstrated to be very common in $\gamma$-ray detected Blazars.

\begin{figure}[ht]
\vbox{
\centerline{
\includegraphics[width=6.5cm, angle=-90] {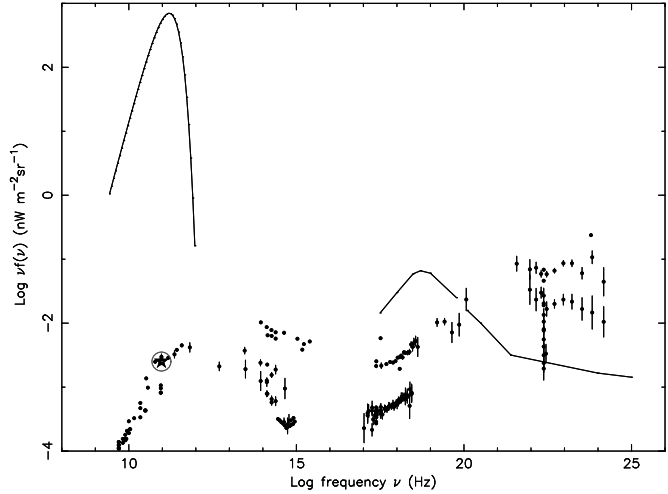}
}}
\caption{The CMB, X-ray and $\gamma$-ray cosmic backgrounds with superimposed the
SED of the Blazar 3C279 scaled so that its flux at 94 GHz matches the cumulative emission
of the entire Blazar population (encircled star symbol).
}
\label{cb_3c279}
\end{figure}

Instead of considering simple average values of the radio to  $\gamma$-ray flux ratio as in \cite{P93}
we estimate the Blazar contribution to the $\gamma$-ray Background using the full SED of Blazars
scaling it to the integrated Blazar flux intensity as calculated with eq. (\ref{eq.IBLazars}). Figure
\ref{cb_3c279} compares the energy distribution of the CMB, CXB and CGB to the SED of 3C 279, a well
known bright EGRET detected Blazar, scaled as explained above.

Figure \ref{cb_3c279} shows the large variability of 3C 279 in the X-ray and $\gamma$-ray band.
While at X-ray frequencies the contribution to the CXB ranges from a few percent to over 10\% in the
higher states, the predicted flux at $\gamma$-ray frequencies ranges from about 100\% to several times
the observed Cosmic Background intensity. This large excess implies that either 3C279 is highly non
representative of the class of Blazars, despite the contribution to the CXB is consistent with other
estimates, or its duty cycle at $\gamma$-ray frequencies is very low (see Fig. \ref{gamma_back}).

The same approach can be followed with other Blazars detected at $\gamma$-ray frequencies.
In most EGRET detected Blazars the SED of LBL Blazars over-predicts the CGB by a large factor.

A way of quantifying the ratio between the  $\gamma$-ray intensity predicted assuming that the source
were representative of the entire population and the actual Background intensity is to use,
in analogy with eq. (\ref{eq.alphamux}), a microwave (94 GHz) to $\gamma$-ray (100 MeV) slope \amg:

\begin{figure}[ht]
\vbox{
\centerline{
\includegraphics[width=7.0cm, angle=-90] {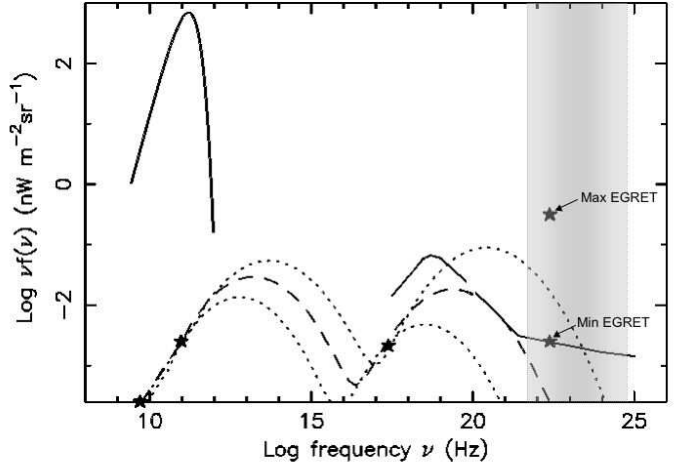}
}}
\caption{The possible contribution of LBL Blazars to the  $\gamma$-ray  Cosmic
Background (shaded area). Simple SSC energy distributions fail to reproduce the observed intensity and slope of the $\gamma$-ray Background. The observed range of  $\gamma$-ray emission with EGRET, normalized to radio flux,  indicates that the duty cycle at these energies must be low (see text for details). }
\label{gamma_back}
\end{figure}

\begin{eqnarray}
 \displaystyle {\alpha_{\mu \gamma}}  & = &
 -{Log(f_{94GHz}/f_{100MeV}) \over {Log(\nu_{94GHz}/\nu_{100MeV})}}
 \nonumber \\
 & = & {Log(f_{94GHz}/f_{100MeV}) \over {11.41}} ~.
 \label{eq.alphamugamma}
\end{eqnarray}

We also define an \amgB as the \amg of an hypothetical source that would produce 100\% of
the CGB if representative of the entire class. This value is based on a CGB intensity
at 100 MeV shown in Fig. \ref{CBGs}. Any real source with \amg flatter than \amgB
would then predict an integrated flux in excess of the observed $\gamma$-ray Background
if representative of the entire population, alternatively its duty cycle must be lower than 100\%.

Table \ref{tab.egret} lists the main properties of the 34 Blazars detected by WMAP and by EGRET.
Column 1 gives the source name, columns 2 and 3 give the Right Ascension and Declination, column 4
gives the 5GHz radio flux, column 5 gives the 94 GHz microwave flux measured by WMAP or estimated from
the extrapolation from measurements in lower WMAP channels or from the literature, column 6 gives the
minimum and maximum $\gamma$-ray flux measured by EGRET (excluding lower limits) taken from the third
EGRET catalog (\cite{Hartman99}), column 7 gives the corresponding \amg values, column 8 gives the
maximum duty cycle allowed assuming that the source is representative of the entire Blazar population,
columns 9 and 10 give the EGRET and WMAP source names. We note that {\it all} sources in the list must
have a duty cycle lower than 100\%.

{\footnotesize
\begin{table*}[h]
\caption{The list and properties of all WMAP-detected Blazars associated to EGRET $\gamma$-ray sources}
\begin{center}
\begin{tabular}{llrccccclc}
\hline \multicolumn{1}{c}{Blazar Name} & \multicolumn{1}{c}{R.A.} & \multicolumn{1}{c}{Dec} &
\multicolumn{1}{c}{Radio flux} & \multicolumn{1}{c}{WMAP flux} & \multicolumn{1}{c}{EGRET flux} &
\multicolumn{1}{c}{\amg} & \multicolumn{1}{c}{Duty} & \multicolumn{1}{c}{EGRET} &
\multicolumn{1}{c}{WMAP } \\
 &J2000.0&J2000.0& 5 GHz & 94 GHz & $>$100 MeV & & cycle & \multicolumn{1}{c}{name}& \multicolumn{1}{c}{catalog}\\
 &         &       &Jy   & Jy  &$10^{-8}$cm$^{-2}$s$^{-1}$ & &\multicolumn{1}{c}{\%} &\multicolumn{1}{c}{3EG J}  &\multicolumn{1}{c}{number}\\
\multicolumn{1}{c}{(1)} &  \multicolumn{1}{c}{(2)} & \multicolumn{1}{c}{(3)} & (4) & (5) & (6) &
(7) & (8) & \multicolumn{1}{c}{(9)} & (10) \\
\hline
 4C15.05      &  02 04 50.3 & 15 14 10  &   3.073   &  1.6$^*$ &9-53 &0.846-0.914 &2-12 & 0204+1458& 092\\
 1Jy0208-512  &  02 10 46.2 &$-$51 01 02 &    3.198 &    1.8 &35-134 &0.816-0.867 &1-4 & 0210-5055&158\\
 B2 0234+28   &  02 37 52.3 & 28 48 08   &  2.794   &   2.1$^*$ &11-31 &0.877-0.917 &5-13 & 0239+2815& 093\\
 CTA26        &  03 39 30.8 &$-$01 46 35 &    3.014 &    3.2 &13-178 &0.827-0.926 &1-17 & 0340-0201& 106\\
 PKS 0420-01  &  04 23 15.7 &$-$01 20 32 &    4.357 &    3.9 &9.3-64.2 &0.873-0.946 &4-29 & 0422-0102& 110\\
 1Jy0454-463  &  04 55 50.7 &$-$46 15 59 &    1.653 &    3.8 &5.5-22.8 &0.911-0.966 &11-47 & 0458-4635& 151\\
 1Jy0454-234  &  04 57 03.1 &$-$23 24 51 &    1.863 &    2.7 &8.1-14.7 &0.915-0.938 &13-23 & 0456-2338& 128\\
 PKS 0506-61  &  05 06 44.0 &$-$61 09 40 &    1.211 &   1.1$^*$ &6-29 &0.855-0.915 &3-13 & 0512-6150& 154\\
 1Jy0537-441  &  05 38 51.3 &$-$44 05 11 &    4.805 &    6.7 &16.5-91.1 &0.880-0.945 &5-28 & 0540-4402& 148\\
 PKS 0735+178 &  07 38 07.3 & 17 42 18   &  1.812   &   1.7$^*$ &15-29 &0.872-0.896 &4-8 & 0737+1721& 113\\
 B2 0827+24   &  08 30 52.0 & 24 10 57   &  0.886   &   2.6$^*$ &16-111 &0.837-0.911 &2-11 & 0829+2413& 112\\
 S50836+710   &  08 41 24.4 & 70 53 40   &  2.342   &   1.2$^*$ &9-33 &0.854-0.903 &2-9 & 0845+7049& 089\\
 OJ 287       &  08 54 48.8 & 20 06 30   &  2.908   &   2.5  &9.7-15.8 &0.910-0.928 &11-18 & 0853+1941& 115\\
 4C 29.45     &  11 59 31.7 & 29 14 43   &  1.461   &   2.1  &7.5-163.2 &0.814-0.931 &1-19 & 1200+2847& 111\\
 PKS1221-82$^a$& 12 24 54.3 &$-$83 13 10 &   0.797   &   1.2$^*$ &11-36 &0.850-0.895 &2-7 & 1249-8330& 178\\
 1Jy1226+023  &  12 29 06.3 & 02 03 04   & 36.923   &    9.0 &8.5-48.3 &0.916-0.982 &13-73 & 1229+0210& 170\\
 3C279        &  12 56 11.0 &$-$05 47 19 &   11.192 &   19.0 &15-250 &0.882-1.000 &5-100 & 1255-0549& 181\\
 PKS 1313-333 &  13 16 07.9 &$-$33 38 59 &    1.093 &   1.3$^*$ &15-32 &0.858-0.887 &3-6 & 1314-3431& 182\\
 1Jy1406-076  &  14 08 56.4 &$-$07 52 25 &    1.080 &   1.7$^*$ &10-128 & 0.815-0.912 &1-12 & 1409-0745& 203\\
 1Jy1424-418  &  14 27 56.2 &$-$42 06 19 &    2.597 &   1.5$^*$ &12-55 &0.842-0.901 &2-9 & 1429-4217& 191\\
 1Jy1510-089  &  15 12 50.4 &$-$09 06 00 &    3.080 &    1.7 &12.6-49.4 &0.851-0.903 &2-9 & 1512-0849& 207\\
 1Jy1606+106  &  16 08 46.0 & 10 29 07   &  1.412   &    3.1 &21.0-62.4 &0.865-0.907 &3-10 & 1608+1055& 009\\
 DA 406       &  16 13 40.9 & 34 12 46   &  2.324   &    1.4 &19-68.9 &0.831-0.880 &1-5 & 1614+3424& 023\\
 4C38.41      &  16 35 15.4 & 38 08 04   &  3.221   &    4.2 &31.8-107.5 &0.856-0.902 &3-9 & 1635+3813& 033\\
 PMNJ1703-6212&  17 03 36.2 &$-$62 12 39 &    0.616 &   1.9$^*$ &14-53 &0.853-0.904 &2-9 & 1659-6251$^b$& 198\\
 S41739+522   &  17 40 36.9 & 52 11 42   &  1.699   &   1.2$^*$ &10-45 &0.842-0.899 &2-8 & 1738+5203& 048\\
 PKS 1814-63$^c$  &  18 19 34.9 &$-$63 45 47 & 4.506 &   1.3$^*$ &14-27 &0.864-0.889 &3-6& 1813-6419& 200\\
 PMNJ1923-2104&  19 23 32.1 &$-$21 04 33 &    2.885 &   2.1$^*$ &29$^{**}$ &0.880 &5 & 1921-2015& 008 \\
 PKS 2052-47  &  20 56 15.5 &$-$47 14 37 &    2.026 &   1.3$^*$ & 9-35 &0.854-0.906 &3-10 & 2055-4716& 208\\
 BL Lac       &  22 02 43.2 & 42 16 39   &  2.940   &   3.8$^*$ &9-40 &0.890-0.947 &7-29 & 2202+4217& 058\\
 PKS2209+236  &  22 12 05.9 & 23 55 39   &  1.123   &   1.3$^*$ & 7-46 &0.844-0.916 &2-13 & 2209+2401& 050\\
 CTA102       &  22 32 36.3 & 11 43 50   &  3.967   &    3.1 &12.1-51.6 &0.873-0.928 &4-18 & 2232+1147& 047\\
 1Jy2251+158  &  22 53 57.6 & 16 08 52   & 14.468   &    5.9 &24.6-116.1 &0.866-0.925 &3-16 & 2254+1601& 055\\
 1Jy2351+456  &  23 54 21.6 & 45 53 03   &  1.127   &   1.7$^*$ &12-43 &0.874-0.923 &4-15 & 2358+4604& 074\\
\hline
\end{tabular}
\label{tab.egret}
\end{center}
$^*$ Not detected by WMAP in the W-band; flux taken from 90GHz measurements \\ $^{**}$ Detected only
once by EGRET reported in NED or extrapolated from lower (K,Ka,Q, V-Band) WMAP channels. \\ $^a$ This
source does not have an optical spectroscopic identification yet. However, the radio SED is very flat,
typical of Blazars. We propose this as a Blazar candidate and associate it with the EGRET source 3EG
J1249-8330 (previously tentatively associated to PMNJ1249-8303 by \cite{mattox97}). \\ $^b$ This EGRET
detection has also been associated to the bright radio source PMN J1647-6437 (\cite{mattox97}) which
is further away from the $\gamma$-ray  position. \\ $^c$ This source is a radio galaxy with steep
radio spectrum at centimeter wavelengths. At microwave frequencies where WMAP is sensitive its
spectrum is flat and indistinguishable from a that of a Blazar.
\end{table*}
}

\subsection{TeV Background}

All Blazars so far detected at TeV energies are of the HBL type. This is readily
interpreted within the SSC scenario since only objects where the synchrotron radiation
extends to near or within the X-ray band can produce a corresponding Inverse Compton flux that
reaches TeV energies, at least assuming a single scattering.

In the following we estimate the Blazar contribution to the TeV background in a graphical way as in
the previous paragraph for the case of the $\gamma$-ray Background but only considering the HBL
component and not the entire Blazar population. From a comparison between the normalization of the
Blazar radio LogN-LogS and that of the Sedentary survey (\cite{Gio99}; see also Fig. \ref{logns} which
is representative of extreme HBLs), we estimate that these objects are about 0.1 \% of the Blazar
population.

\begin{figure}[ht]
\vbox{
\centerline{
\includegraphics[width=6.5cm, angle=-90]{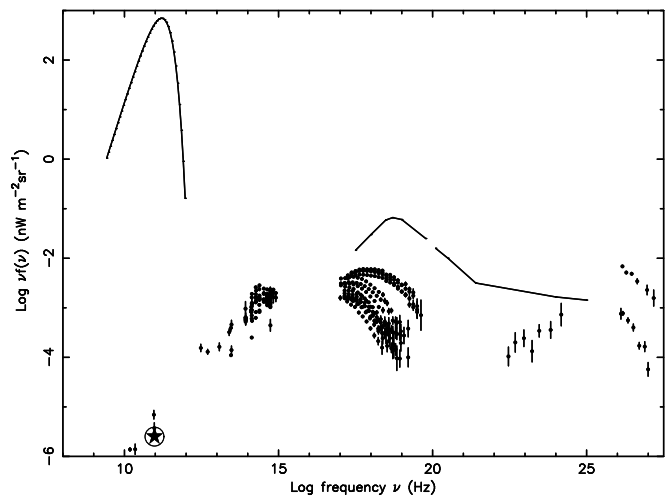}
}}
\caption{The CMB, X-ray and $\gamma$-ray cosmic backgrounds with superimposed the
SED of MKN 421 scaled to 1/1000 of the total Blazar contribution to the CMB since
the radio LogN-LogS of extreme HBL like this source (in the sedentary survey) is
about 1/1000 of the LogN-LogS of all Blazars.
}
\label{cb_mkn421}
\end{figure}

In Fig. \ref{cb_mkn421} we plot the SED of the well known TeV detected Blazar MKN421 normalized at 94 GHz so that the flux is scaled to 0.1\% of the intensity produced by the entire population of Blazars.

From this figure we see that, despite HBLs are a tiny minority, their integrated X-ray flux makes
a fair fraction of the CXB and that their TeV emission may produce a significant amount of extragalactic
light, even taking into account that the real flux level should be lower than that shown in Fig. \ref{cb_mkn421} at TeV energies since only the objects closer than z$\approx 0.2 $ can be detected at these energies.

We note however that since extreme HBLs such as those of the Sedentary survey are
very rare (one object in several thousand degrees with flux above a few mJy) the
extragalactic light at TeV energies should be very patchy, associated to single sources,
rather than a diffuse light resulting from the superposition of many unresolved discrete sources
as in the other cosmic backgrounds.

\section{Summary and discussion}

Blazars are the only type of AGN known to emit non-thermal radiation across the entire electromagnetic spectrum, from radio waves to the most energetic $\gamma$-rays.
In some energy bands the power emitted by these sources is orders of
magnitude larger than that generated through the accretion process by other types of AGN.
At these frequencies, despite their low space density, Blazars are the dominant population in the extragalactic sky and contribute in a significant way to the extragalactic Cosmic Backgrounds.

A deep understanding of the Blazar contribution to the Cosmic Background light is becoming an increasing necessity as the microwave, the $\gamma$-ray and the TeV band are about to be intensively explored by a new generation of astronomy satellites and ground-based Cherenkov telescopes.

The overall Cosmic Background energy has two well understood components: the primordial Black Body emission peaking at microwave frequencies, or Cosmic Microwave Background (CMB) and the X-ray apparently diffuse emission arising from the accretion onto super-massive black holes in AGN integrated over cosmic time, or Cosmic X-ray Background (CXB). We have shown that Blazars
add a third non-thermal component that at low frequencies contaminates the CMB fluctuation spectrum and complicates its interpretation, while at the opposite end of the electromagnetic spectrum dominates the extragalactic background radiation.

Our calculations are based on a new, deep Blazar radio LogN-LogS that was assembled combining several radio and multi-frequency surveys. The integrated radio flux from the entire population has been extrapolated to other energy bands using observed flux ratios and broad-band Spectral Energy Distributions.

Our results can be summarized as follows.\\

\noindent {\bf Contribution to the CMB}. The results presented here confirm and extend the findings of
\cite{giocol04}. The Blazar contribution to the Cosmic Background intensity at microwave frequencies
ranges between $5\times 10^{-5}$ to $5\times 10^{-6}$ of the CMB, depending on observing wavelength.
The consequences are twofold: i) there is an apparent temperature excess of 5-50 $\mu $K, and ii) the
CMB fluctuation spectrum is affected by a spurious signal that becomes significant or dominant at
multipole $l \approx 300-600,$ (see Figs. \ref{cmb41} and \ref{cmb94}). Since Blazars, as all cosmic
sources, are probably not distributed in a completely random way across the sky, the source clustering
may significantly increase the amount of contamination, particularly at large angular scales
(\cite{Gonzales05}).

We also note that the temperature excess causes a bias in the statistical distribution of the
primordial CMB fluctuation spectrum. If not properly removed, this non-thermal foreground radiation
may greatly complicate the detection of primordial non-Gaussianity that may carry important
information on the dynamics of  the inflationary phase (e.g., \cite{Peebles97}).\\

\noindent {\bf Contribution to the CXB.} The contribution to the soft X-ray Background has been
derived converting the integrated radio and microwave Blazar emission into X-ray flux using observed
flux ratio distributions.  Our estimated values of  11-12\% of the CXB are in very good agreement with
the independent measurement of  13\% for radio loud AGN obtained by  \cite{Galbiati04} using
XMM-Newton data. For the large part this X-ray flux is composed of emission due to the end of the
synchrotron component of HBL objects, while for about one third the X-ray flux is due to the flat
Inverse Compton spectral component of LBL sources. Given the very different X-ray spectral slopes of
LBL and HBL Blazars,  at energies higher than 1 keV, the blend between the (steep) synchrotron HBL
component and the (flat) Inverse Compton LBL component must change in favour of the latter. Since the
spectral slope of the Inverse Compton component is very similar to that of the CXB up to 40-50 keV,
the LBL contribution should stay roughly constant around 4-5\% up to that energy. Above $\approx $  50
keV the observed CXB steepens, and the contribution of LBL Blazars  should progressively increase. One
of the methods that we have used is based on the \fxfr distribution of Fig. \ref{fxfr} where LBL
sources are the large majority among Blazars while HBL Blazars are only about 5\%. We have converted
the integrated radio flux into X-rays assuming  that  the \fxfr distribution remains  constant at {\it
all radio fluxes}. A strong increase of the fraction of HBL Blazars at low radio luminosity (hence
fluxes), as required by the Blazar sequence within the unified schemes of \cite{Fos97} and
\cite{Ghi98}, would result in a much larger Blazar contribution to the CXB, inconsistent with the
observations. This conclusion, together with the findings of \cite{Gio02b}, \cite{P03}, \cite{Cac04},
\cite{Gio05} and \cite{Nieppola05} cast serious doubts on the validity of the Blazar sequence.\\

\noindent {\bf Contribution to the soft $\gamma$-ray Background.} At energies higher than $\approx $100 keV the contribution to the Cosmic Background flux was estimated converting the expected Blazar contribution to the diffuse background at microwave frequencies using SEDs predicted by simple homogeneous Synchro-Self-Compton models,
constrained to have a microwave-to-X-ray spectral slope \amx equal to the observed average
value of 1.07.

The Hard X-ray/Soft $\gamma$-ray Background is reproduced both in shape (spectral slope) and intensity assuming an average SSC distribution where the synchrotron component peaks
at $10^{13.5}$ Hz,  well within the range of observed values.
We conclude that the Cosmic Background between $\approx $0.5 and $\approx $ 10 MeV is consistent with being due to the tail of the  Inverse Compton component of LBL objects (see Fig. \ref{hardx_back}) . \\

\noindent {\bf Contribution to the $\gamma$-ray Background.}
The SED of Blazars detected  by EGRET predict much more  $\gamma$-ray Background
than observed and therefore they cannot be representative of the entire population, at least in a stationary situation. Either EGRET detected sources are special, non representative objects, or their
$\gamma$-ray duty cycle must be low. Indeed, strong variability  at  $\gamma$-ray energies is very common and
most objects have been detected in widely different intensity states.  A plausible scenario for the
origin of  the $\gamma$-ray Background is that it is due to a mixture of Inverse Compton radiation produced by LBL Blazars during strong flares (Fig. \ref{gamma_back}) and perhaps a less variable component  due to the still rising part of the Compton spectrum in HBL objects (Fig. \ref{cb_mkn421}). \\

\noindent {\bf A TeV cosmic Background?} The existence of a Cosmic Background at TeV energies has not yet been established.  Blazars of the HBL type, especially those where the synchrotron peak is located at very high energies can in principle produce a significant integrated flux (Fig. \ref{cb_mkn421}). A precise prediction is however difficult  given the uncertainties in the amount of absorption of TeV photons via interaction with photons of the Infra-red background. In addition, we note that since the space density of extreme HBL is very low  (less than one object in one hundred square degrees at a radio flux of 3.5 mJy) the TeV background  produced by Blazars is composed of widely separated discrete sources rather than an apparently diffuse light like e.g. the Cosmic X-ray Background.  \\

In all our estimations we have assumed that the broad band SED of Blazars is characterized by
approximately equal power in the syncrotron and inverse Compton components.  We note that an average
SED, where the inverse Compton component is much more powerful than the synchrotron component, is not
acceptable as it would predict a much larger than observed background at hard-X-ray/soft Gamma ray
energies (see Fig. \ref{hardx_back}). The large inverse Compton emission compared to synchrotron flux
observed in some LBL sources (e.g., \cite{Ballo02}) should therefore be associated to either rare
objects or transient events rather than to an average emission.\\

\begin{figure}[ht]
\vbox{
\centerline{
\includegraphics[width=8.0truecm, angle=-90] {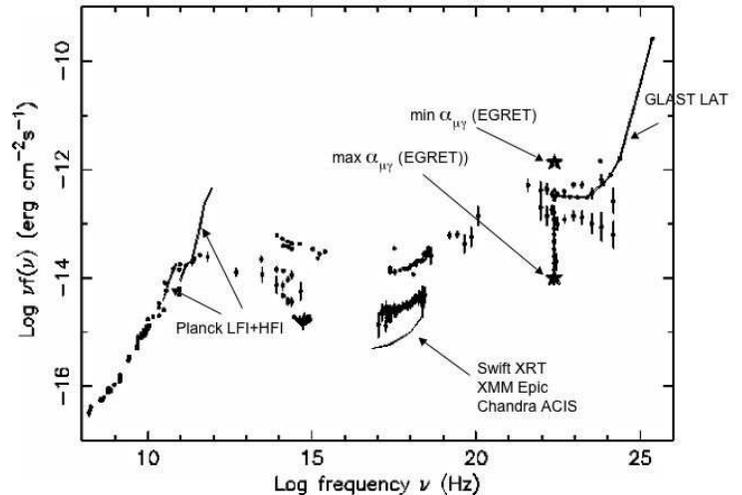}
}}
\caption{The SED of the Blazar 3C279 scaled down by a factor of 1000 (1 milli-3C279) with superimposed the
sensitivity of microwave (Planck LFI+HFI), X-ray (Swift XRT-XMM-Chandra) and $\gamma$-ray (GLAST-LAT) satellites.
A Blazar of this flux, or about 10 mJy at radio frequencies, would still be detected by Planck,
by existing optical and X-ray satellites (Swift, XMM or Chandra) and would be above the GLAST limiting sensitivity during strong flares.
}
\label{m3c279}
\end{figure}

\noindent {\bf Blazars and future observatories.} Figure \ref{m3c279} shows the observed SED of the
well known LBL Blazar 3C279, scaled down by a factor 1000, overlaid with the limiting sensitivities of
the upcoming Planck and GLAST  satellites and with a 0.5-10 keV sensitivity that can be reached by
existing X-ray observatories. This hypothetical $\sim 10$ mJy LBL Blazar (or 1 milli-3C279) is at the
limit of the Planck sensitivity, it is detectable with a deep Swift exposure (or a less deep
XMM-Newton or Chandra observation) and it is detectable by GLAST during strong flares. Since the radio
LogN-LogS of Fig. \ref{logns} predicts a space density of $\gsim 5$ objects per square degree with
flux above 10 mJy, the Planck mission should detect $\approx $ 100,000-200,000 Blazars  in the
$\approx 30,000$ square degree high galactic latitude sky. A fraction of these sources will also be
detected by GLAST when flaring, exactly how many strongly depends on the duty cycle.    \\

\noindent {\bf A deep all-sky X-ray survey to clean the data from Planck and future CMB missions and to provide a database of $>$ 100,000 Blazars for GLAST and other $\gamma$-ray observatories.}
Given the significant impact of the Blazar foreground emission on the CMB power spectrum
it is important to remove as much as possible this contaminating component from the CMB.
One possible efficient way to achieve this is to exploit the fact that the spectral slope distribution between microwave and soft X-ray flux of LBL Blazars is very narrow (see Fig. \ref{a94x}) with a dispersion that is probably mostly due to intrinsic variability. The soft X-ray flux of LBL Blazars (that is $>$90\% of the Blazar population) is therefore a very good estimator of the flux at microwave frequencies and could be used to locate and remove foreground Blazars from the CMB.  Figures \ref{a94x} and \ref{m3c279}  imply that
a hypothetical all-sky survey with limiting sensitivity of a few $10^{-15}$ \ergs in the soft X-ray band, would detect the large majority of  Blazars above the limiting sensitivity of Planck and therefore allow the construction of a database including well over 100,000 Blazars with flux measurements at radio, microwave and X-ray frequencies.
Although this type of surveys are currently not planned for the near future, they are clearly needed to address the previous issues. The extremely large sample of Blazars  produced by such survey could also be used
to study with great detail the statistical properties of Blazars, including the spatial correlation function and would provide a large number of targets for the next generation of $\gamma$-ray observatories such as AGILE, GLAST and future instruments operating in the still poorly explored MeV spectral region.

\begin{acknowledgements}

This work is partly based on XMM-Newton, \sax and ROSAT X-ray archival data taken from the the ASI Science Data Center (ASDC),  Frascati, Italy, and on data taken from the following on-line services \\
the NASA/IPAC Extragalactic Database (NED) and the Sloan Digital Sky Survey (SDSS, Data Release 3).  \\
 We are grateful to  Paolo Padovani for providing DXRBs LogN-LogS results in advance of publication and to
 Enrico Massaro for useful discussions.\\ S.C. is supported by PRIN-MIUR under contract No.2004027755$\_$003.
\end{acknowledgements}


\end{document}